\documentclass[11pt,twoside]{article}
\usepackage{asp2010_JCB}

\resetcounters

\usepackage{graphicx}

\begin{document}

\addtocounter{page}{235}

\title{Modeling the Galactic Magnetic Field using Rotation Measure Observations in the Galactic Disk from the CGPS, SGPS, and the VLA} 
\author{Cameron Van Eck and Jo-Anne Brown  
\affil{Department of  Physics and Astronomy, University of Calgary, Canada}
}   

\begin{abstract}

Interstellar magnetic fields play critical roles in many astrophysical processes. Yet despite their importance, our knowledge about magnetic fields in our Galaxy remains limited. For the field within the Milky Way,  much of what we do know comes from observations of polarisation and Faraday rotation measures (RMs) of extragalactic sources and pulsars. A high angular density of RM measurements in several critical areas of the Galaxy is needed to clarify the Galactic magnetic field structure. 
Using observations made with the VLA, we have determined RMs for sources in regions of the Galactic plane not covered by the Canadian Galactic Plane Survey (CGPS) and Southern Galactic Plane Survey (SGPS).  We have combined these new RMs with those determined from the CGPS and SGPS and have produced a new model for the magnetic field of the Galactic disk.  
\end{abstract}

\section{Introduction}

The Galactic magnetic field plays critical roles in the interstellar medium, ranging from star formation
to large-scale galactic dynamics.  However, much remains unknown
about how the field is generated or how it is evolving.  The only way to make progress in addressing these questions is to 
fully understand the present overall structure of the field.  This is an essential constraint to proposed evolutionary models of the field.

One observation essential to the study of the Galactic magnetism is that of the 
rotation measure (RM).
As a linearly polarised electromagnetic wave propagates through a region containing free thermal electrons and
a magnetic field, such as the interstellar medium,  the plane of polarisation will rotate through the
process known as Faraday rotation.    If we assume the polarised radiation emitted by a source is at a constant 
angle, $\tau_\circ$, and that this radiation is {\it{only}} affected by Faraday rotation, then the polarisation angle that 
we measure, $\tau$, will be given by
\begin{equation}
\tau = \tau_\circ + 0.812 \lambda^2 \int_{\rm source}^{\rm receiver} n_e \mathbf{B \cdot} {\rm{d}}\mathbf{l} =\tau_\circ+ \lambda^2 {\rm RM},
\end{equation}
where  $\lambda$ is the wavelength, $n_e$[cm$^{-3}$] is the electron density, $\bf{B}[\mu$G] is the magnetic field, d$\bf{l}$[pc] is the path length element.  
Consequently, measuring the polarisation angle at several wavelengths for a given source can provide a simple determination of the rotation measure for that line of sight.
Compact polarised radio sources within the Galaxy (pulsars) and outside the
Galaxy (extragalactic sources or EGS) subsequently act as line-of-sight probes for the magnetic field;
the higher the projected spatial density of the  observed probes, the easier it is to determine the
field structure in the given region.

Recent work has focused on developing and testing competing models and on determining the existence of large scale reversals in the magnetic field.  Magnetic field reversals occur where the magnetic field direction completely reverses over a short change in radius and/or azimuth  within the disk of the Galaxy.  The number of reversals depends on the interpretation of the existing RM data and is presently a very controversial subject \citep[e.g.][]{bt01, Weisberg04, Han06,Brown07, Sun08, Vallee08, Men08, Ronnie, Katgert}.  Most models are made to follow the spiral arm structure of the Galaxy since an approximate alignment of the regular magnetic fields and spiral arms is commonly observed
in external galaxies \citep[e.g.][]{Fletcher}. For all of these models, sufficient numbers of low-latitude, high quality RM data in key regions have been lacking.

\section{Observations}

\begin{figure}[htb]
\includegraphics[scale=0.4]{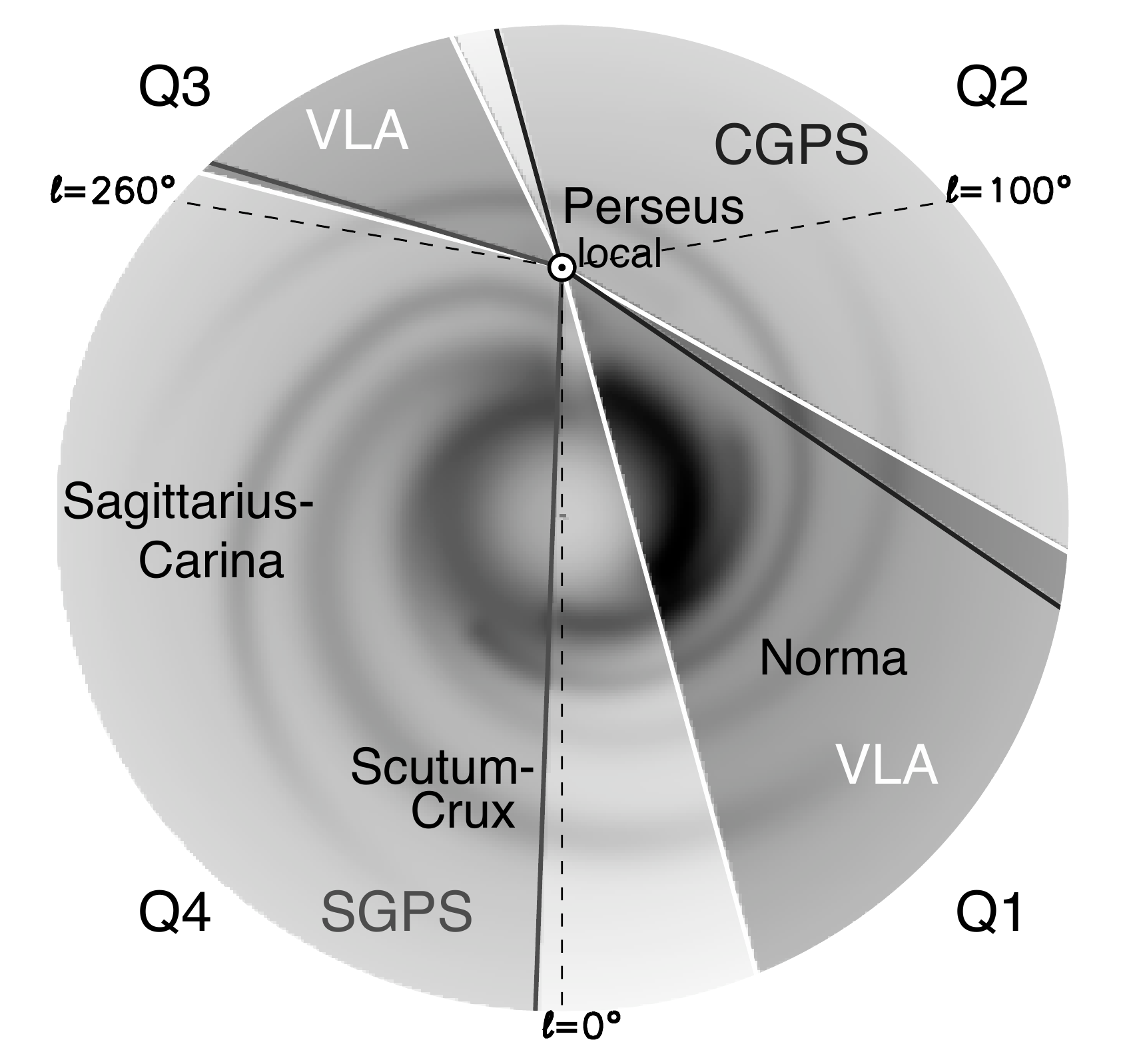}   
\centering
\caption{View of the Milky Way from above the north Galactic pole illustrating the main survey regions of extragalactic rotation measures used in this paper.  The grey scale background is the electron density distribution model of Cordes and Lazio (2002).  The dark lines are the boundaries of the regions observed by the CGPS and SGPS, while the white lines denote the 2 areas targeted by the VLA data used for this project.  The dashed lines show the delineations between the three sections we modeled.} 
\label{fig: Coverage}
\end{figure}

Two recent projects have produced catalogues for several hundred EGS RMs in the plane of the Galaxy: the Canadian Galactic Plane Survey \citep[CGPS;][]{CGPS, CGPSRMs} and the Southern Galactic Plane Survey \citep[SGPS;][]{Haverkorn06, Brown07}.  These surveys left two gaps in the EGS RM coverage of the Galactic plane as shown in Figure \ref{fig: Coverage}. Using the Very Large Array (VLA), we performed observations to fill in the gaps in RMs for EGS between the CGPS and SGPS.

\begin{figure}[htb]
\centering
\includegraphics[scale=0.51]{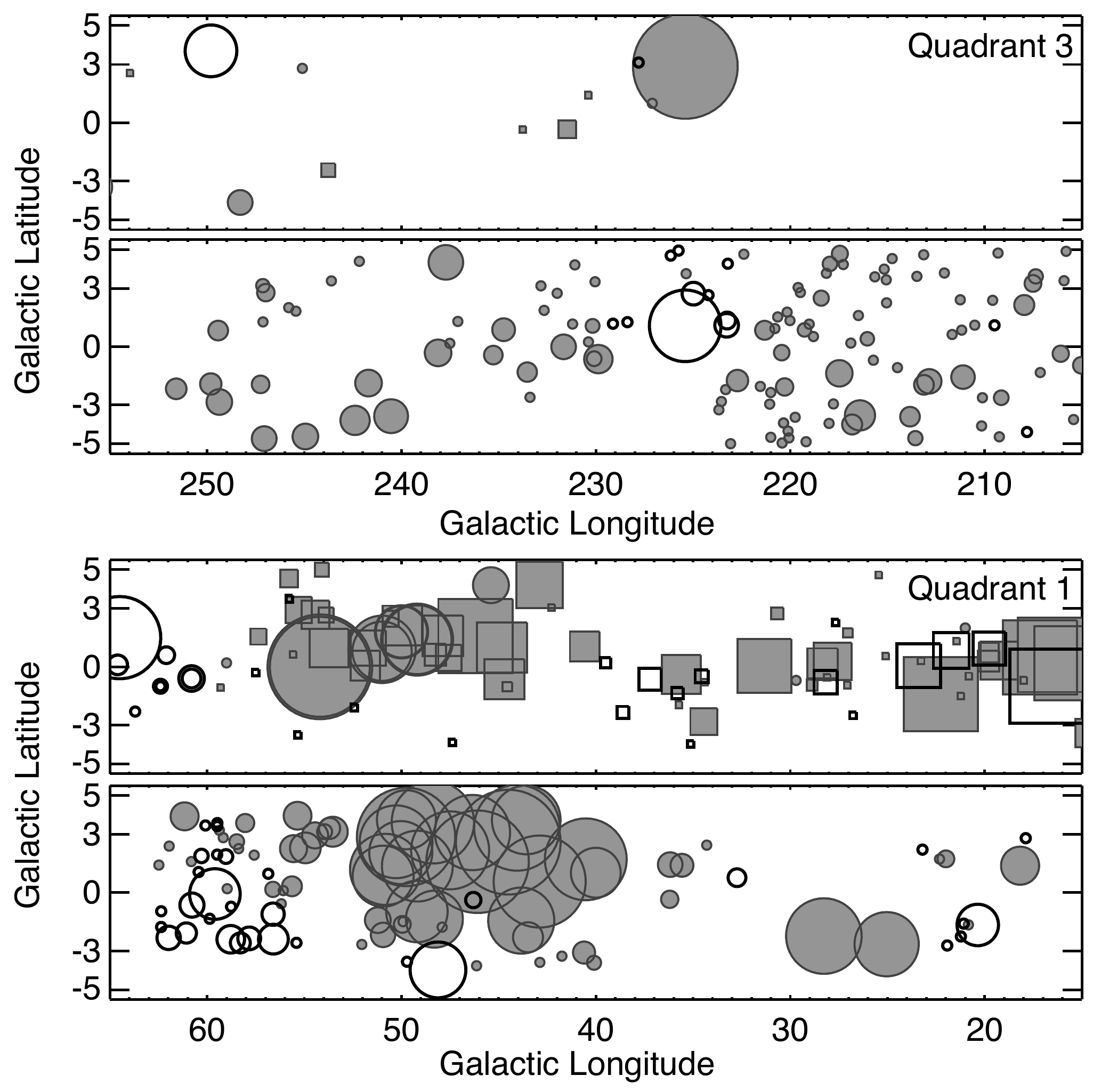}  
\caption{A comparison of RMs in the VLA regions before and after the Van Eck et al. (2010) catalog.  The top panels for each quadrant (1 and 3) are previous EGS (circles) and pulsar (squares) RMs.  The bottom panels for each quadrant are the new EGS RMs (circles).   Grey filled symbols represent positive rotation measures; black open
symbols represent negative rotation measures; diameters of symbols are linearly scaled to the magnitude of  RM truncated between
 100 and 600 rad m$^{-2}$ so that sources with $|$RM$| < $ 100 rad m$^{-2}$ are set to 100 rad m$^{-2}$, and those with
 $|$RM$| > $ 600 rad m$^{-2}$ are set to 600 rad m$^{-2}$.} 
\label{fig: Circle_plots}
\end{figure}

From the VLA observations, we determined  reliable RMs for 221 EGS, 96 of which are in quadrant 1 and 125 in quadrant 3.  For details on the processing and the RM catalog, see \citet{VanEck10}.  These RMs are shown in Figure \ref{fig: Circle_plots}, in comparison with the previously published RMs in these regions.

\section{Multi-sector Model of the Magnetic Field in the Galactic Disk}

 Many recent works  have focused on building an empirical model of the magnetic
 field for the whole Galaxy \citep[e.g.][]{Weisberg04,Han06, Vallee08}.  By
 contrast, we have
 chosen to take a `hybrid' approach to our modeling work.  We examine the entire
 Galactic disk, but in 3 separate sectors to see if we can determine any
 common features or structure across the sectors. We purposely do not apply any `boundary
 matching' conditions between the sectors in order to facilitate independent
 results for each of the three sectors examined.  It was our intention to see
 if there was any commonality amongst the different sectors that could be
 arrived at independently.  The three sectors we examine are as follows:  the
 outer Galaxy, defined roughly as spanning quadrants 2 and 3 (Q2 and Q3), and
 the two inner Galaxy regions, roughly defined by quadrants 1 and 4 (Q1 and
 Q4).

We use the method described by \citet{Brown07} and in greater detail in
 \citet{joannethesis}.  In summary, we attempt to empirically reproduce the
 observed RMs of both pulsars and EGS using the electron density of
 \citet[][hereafter NE2001]{NE2001}, and various magnetic field models. With the
goal of our modeling being to explore the large-scale field, we ignore the
small-scale clumps and voids of NE2001, and use only the thin, thick, and
spiral arm components.

We also placed the following restrictions on all of the models we investigated. First, the magnetic field
for Galactocentric radii
$R > 20$ kpc or $R<3$
kpc was set to zero. Similarly, the field was assumed to be
zero for
$|z| > 1.5$ kpc.
  In addition, all models have a circular region containing the molecular ring
of the NE2001 electron model (3 kpc $\le R \le 5$ kpc)
with a circular magnetic field regardless of the geometry being tested in the rest of
the Galaxy.

We used EGS RMs our VLA observations discussed here,  as well as the SGPS and CGPS
data sets.   We treated the EGSs as being located at the edge of the Galaxy
(modeled as $R=20$ kpc).  Since we did not  wish to investigate the complicated nature of the
field likely to be found near the Galactic center,
we did not use
any EGS RM sources within $\pm 10^\circ$ Galactic center. 
Since we did not consider
any vertical structure in our model, we removed any sources with a
calculated height
$|z|>1.5$ kpc.   With these criteria, we were left with
211 of the 221 RMs described in section 2, 142 of the 148 RMs from the
SGPS and 1020 sources from the CGPS,
a total of 1373 EGS
sources.

We used 557 pulsar RMs from the following sources:
\citet{Noutsos}, \citet{Han06}, \citet{Weisberg04}, \citet{mitra}, \citet{Han99}
and \citet{pulsars}  where pulsars were selected with
$|z| < 1.5 $ kpc.  For
self-consistency,
we used distances to the pulsars
predicted by the NE2001 model.

The EGS and pulsar RMs were then split into the sectors described in the next section.
Using the resulting best-fit values for the magnetic field model, we could then 
calculate `predicted' RMs and compare them to the observed RMs.  Through this
comparison, we could assess the quality of fit and vary the model in each sector 
to test for different attributes of the magnetic field. 
Below we present what we found to be our   best  model in each sector.  Additional 
details of our modeling and the assessment of the quality of fit is given in \citet{VanEck10}.

\subsection{Model Sector I: The Outer Galaxy (Q2 and Q3)}

This sector is strictly the outer Galaxy, which we define  to be
$100^\circ < \ell < 260^\circ$.   Based on previous experience, we expected this region to be
the simplest in nature.  Within this region, 
 we have RMs for 88 pulsars, and 864 EGS (21 from SGPS, 125 VLA, 718 from CGPS).

As demonstrated by earlier CGPS work, RM data in the outer Galaxy hold 
no strong evidence for
a reversal \citep{bt01, joannethesis,
Brown03}.   There is some evidence that suggests that
the field decays as $1/R$, consistent with the decay in electron density \citep{Brown03}.  Therefore, we
modeled this region as a single magnetic entity with a $1/R$
 decay in magnitude. Coupled with the results of Rae et al. (these proceedings), we found  a purely
azimuthal field to fit the best in this region.

 \subsection{Model Sector II: SGPS region (Q4)}

 We define this region to be  $260^\circ < \ell < 360^\circ$, which is slightly less than the area modeled by \citet{Brown07}.
This region contains 292 pulsars and 121 EGS (all from the SGPS).

 For their model, \citet{Brown07} used the available pulsar RMs combined with
new EGS RMs from the SGPS to model the magnetic field within the SGPS region. 
 Given how well the model agreed with the data, we decided to keep much of this
 model the same, but revise it slightly to agree with the findings of our
 model sector I.  To that end, we merged all separate regions beyond the
 Sagittarius-Carina arm into one magnetic field region, and redefined the
 field in this new outer-Galaxy region to be purely azimuthal, and
 retained the  $1/R$ dependence in this region, consistent with \citet{Brown03} and \citet{Brown07}.
 However, for the remaining regions in the inner Galaxy we reverted to a
 constant field strength as suggested by \citet{Katgert}.

The best-fit magnetic field results for our variation on the
 \citet{Brown07} model produced predicted RMs that are virtually indistinguishable from that of the original model.  
 However, this new model is less complicated as it has 8 regions compared to 9 regions in
the original model, and has removed the complexity of the
$1/R$ dependence in the inner Galaxy. The reduction in
the number of parameters while maintaining a good quality
of fit, lends support to this model.

 \subsection{Model Sector III: VLA region  (Q1) }

We define this region to be  $0^\circ < \ell < 100^\circ$.  In this region, we have 177 pulsar RMs and 388 EGS RMs (302 from CGPS, 86 from our VLA observations).

For this sector, we used as our starting position the ASS+RING
model proposed by \citet{Sun08}. In particular, we began by  assuming that in
the inner Galaxy, the magnetic region delineations are circular, but the
fields within the regions are spiral.  We again carried over the circular, 1/R outer Galaxy
field from sector I.

For our model, the first region boundary is located at
$R=5.0$
 kpc to correspond to
the molecular ring, and to be consistent with the Q4 model.  The remaining
boundary locations were optimized based on a comparison between the 
predicted RMs and the modeled RMs;
they were found to
be $R=5.7$ kpc, $R=7.2$ kpc, and $R=8.4$ kpc.

\begin{figure}[htbp]
\centering
\includegraphics[scale=0.4]{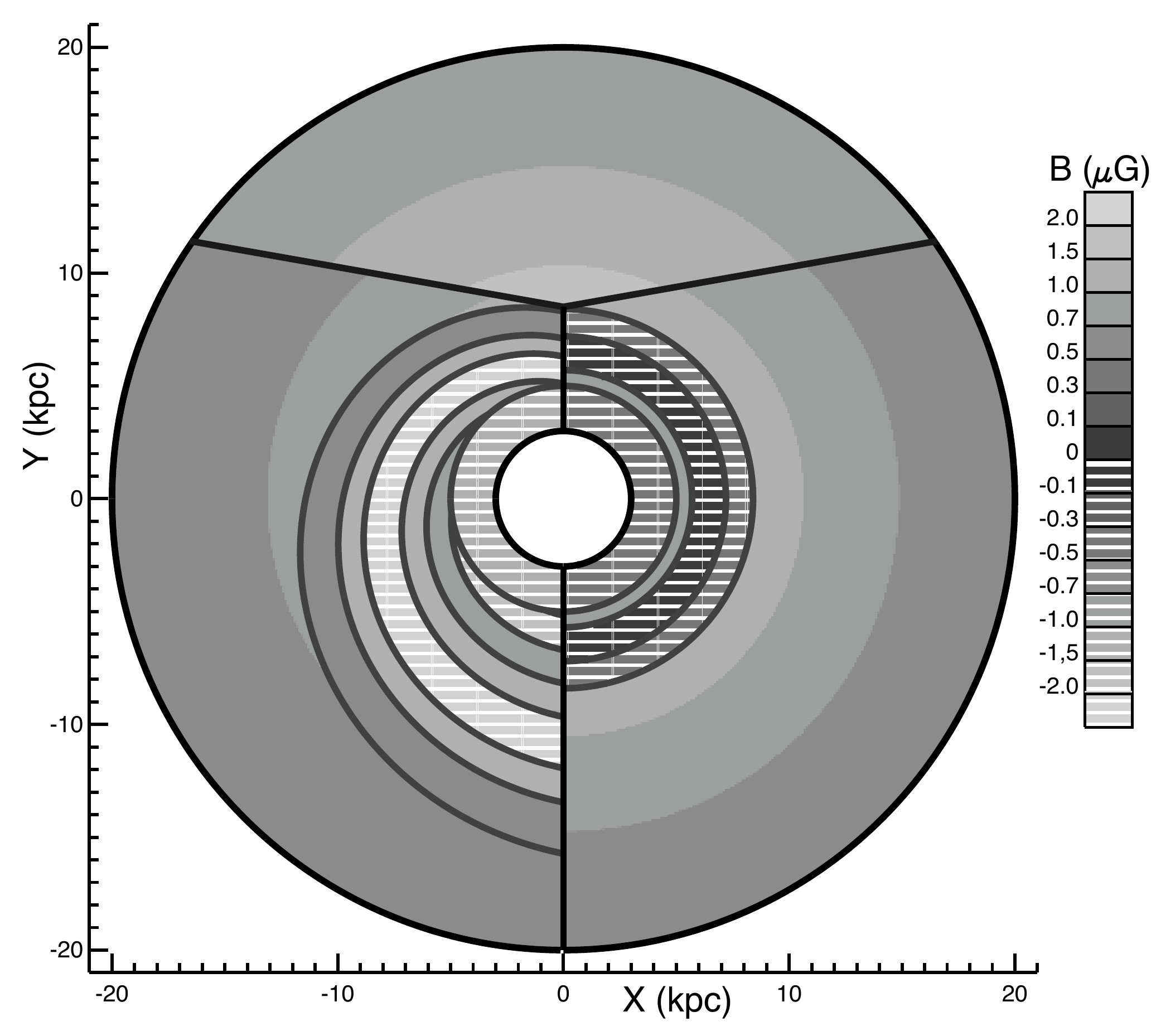}
\caption{Best-fit magnetic field strengths for each of the regions.  Solid shades represent clockwise field, while striped shades represent a counter-clockwise field.} 
\label{fig: Bmap}
\end{figure}

For this region, our best fit magnetic field model has the inner Galaxy primarily oriented counter-clockwise, except for
one small clockwise piece next to the molecular ring, as shown in Figure \ref{fig: Bmap}.

\subsection{Combining the Sectors}

When we consider our three sectors together, as shown in Figure
\ref{fig: Bmap}, a picture emerges of a predominantly clockwise Galactic
magnetic field with what could be interpreted as single reversed
(counter-clockwise) region spiraling out from the Galactic center.
According to our analysis, the field in the inner Galaxy
has a spiral shape (with a pitch angle estimated here as $11.5^\circ$) and is
generally aligned with the spiral arms while in
the outer Galaxy it is (almost) azimuthal.  Our model may be considered as something of a zeroth-order approximation; it was constructed in a piece-wise manner, yet there is some consistency across the whole of the
Galaxy. We also note that this `spiraling-out'  reversed region could extend
into Q1 at larger Galactic radii, but without any pulsars located on the far
side of the Galaxy in this region, determining the existence of such a region
is not possible.

We expect that significant improvements on this
model, using the same technique and the present edition of the electron
density model, will be difficult to accomplish for several reasons.  First,
the electron density includes very little small scale structure beyond the
local regions.   Second, the reliability of distances to the pulsars remains
questionable; small shifts in the assumed position of the pulsars will
influence the results of the best fit.  Finally, as is always the case with
modeling, more data would vastly improve the model.  For example, pulsars on
the far side of the Galactic center would provide much needed constraints in
the area, and as discussed above, the EGS source density is considerably lower
in this part of the inner Galaxy as well. However, such data are unlikely to
be available
until higher sensitivity instruments, such as ASKAP or the SKA
come online.

\section{Summary and Discussion}

We have  produced a catalog of 221
rotation measures of extragalactic sources to fill in
critical gaps in the disk rotation measure coverage of the Canadian Galactic
Plane Survey and the Southern Galactic Plane Survey.
We then used these data, in conjunction with previously observed rotation measures, to propose a new magnetic field model, stemming from a new modeling strategy that studies
the disk field in three different sectors. The division of sectors is  roughly
between the outer Galaxy (quadrants 2 and 3), and the two `inner Galaxy' quadrants:  quadrant 1 and quadrant 4.  
Our modeling suggests
that the outer Galaxy is dominated by
an (almost)
purely azimuthal field, whereas
in  the inner Galaxy
magnetic lines have a spiral shape,
and are likely
to be aligned with the spiral arms.
Furthermore, the model seems to indicate that the magnetic field in the Galaxy
is predominantly clockwise, with a single reversed region that appears to
spiral out from the center of the Galaxy.

Additional observations, and more modeling work are needed to confirm our results.  However, we believe that this multi-sector approach to
empirical modeling yields a more reasonable model than attempting to model the entire Galaxy with the current status of
both the electron density and available data.

\acknowledgements 
We would like to thank Rick Perley and the other staff at the NRAO for their assistance in collecting and processing the VLA data.
The National Radio Astronomy Observatory is a facility of the National Science Foundation operated under cooperative agreement by Associated Universities, Inc.
This work was supported in part by grants to J.C.B. and C.V.E. from the Natural Sciences and Engineering Research Council of Canada,
and a grant to C.V.E. from Alberta Innovates - Technology Futures.

\bibliography{VanEck_Cameron}

\begin{thebibliography}{}
\expandafter\ifx\csname natexlab\endcsname\relax\def\natexlab#1{#1}\fi
\expandafter\ifx\csname url\endcsname\relax
  \def\url#1{\texttt{#1}}\fi
\expandafter\ifx\csname urlprefix\endcsname\relax\def\urlprefix{URL }\fi
\providecommand{\eprint}[2][]{\url{#2}}

\bibitem[{{Brown}(2002)}]{joannethesis}
{Brown}, J.~C. 2002, Ph.D. thesis, University of Calgary (Canada)

\bibitem[{{Brown} et~al.(2007){Brown}, {Haverkorn}, {Gaensler}, {Taylor},
  {Bizunok}, {McClure-Griffiths}, {Dickey}, \& {Green}}]{Brown07}
{Brown}, J.~C., {Haverkorn}, M., {Gaensler}, B.~M., {Taylor}, A.~R., {Bizunok},
  N.~S., {McClure-Griffiths}, N.~M., {Dickey}, J.~M., \& {Green}, A.~J. 2007,
  \apj, 663, 258. \eprint{arXiv:0704.0458}

\bibitem[{{Brown} \& {Taylor}(2001)}]{bt01}
{Brown}, J.~C., \& {Taylor}, A.~R. 2001, \apjl, 563, L31

\bibitem[{{Brown} et~al.(2003{\natexlab{a}}){Brown}, {Taylor}, \&
  {Jackel}}]{CGPSRMs}
{Brown}, J.~C., {Taylor}, A.~R., \& {Jackel}, B.~J. 2003{\natexlab{a}}, \apjs,
  145, 213

\bibitem[{{Brown} et~al.(2003{\natexlab{b}}){Brown}, {Taylor}, {Wielebinski},
  \& {Mueller}}]{Brown03}
{Brown}, J.~C., {Taylor}, A.~R., {Wielebinski}, R., \& {Mueller}, P.
  2003{\natexlab{b}}, \apjl, 592, L29

\bibitem[{{Cordes} \& {Lazio}(2002)}]{NE2001}
{Cordes}, J.~M., \& {Lazio}, T.~J.~W. 2002, ArXiv Astrophysics e-prints.
  \eprint{astro-ph/0207156}

\bibitem[{{Han} et~al.(2006){Han}, {Manchester}, {Lyne}, {Qiao}, \& {van
  Straten}}]{Han06}
{Han}, J.~L., {Manchester}, R.~N., {Lyne}, A.~G., {Qiao}, G.~J., \& {van
  Straten}, W. 2006, \apj, 642, 868. \eprint{arXiv:astro-ph/0601357}

\bibitem[{{Han} et~al.(1999){Han}, {Manchester}, \& {Qiao}}]{Han99}
{Han}, J.~L., {Manchester}, R.~N., \& {Qiao}, G.~J. 1999, \mnras, 306, 371.
  \eprint{arXiv:astro-ph/9903101}

\bibitem[{{Haverkorn} et~al.(2006){Haverkorn}, {Gaensler}, {McClure-Griffiths},
  {Dickey}, \& {Green}}]{Haverkorn06}
{Haverkorn}, M., {Gaensler}, B.~M., {McClure-Griffiths}, N.~M., {Dickey},
  J.~M., \& {Green}, A.~J. 2006, \apjs, 167, 230.
  \eprint{arXiv:astro-ph/0609010}

\bibitem[{{Jansson} et~al.(2009){Jansson}, {Farrar}, {Waelkens}, \&
  {En{\ss}lin}}]{Ronnie}
{Jansson}, R., {Farrar}, G.~R., {Waelkens}, A.~H., \& {En{\ss}lin}, T.~A. 2009,
  Journal of Cosmology and Astro-Particle Physics, 7, 21. \eprint{0905.2228}

\bibitem[{{Men} et~al.(2008){Men}, {Ferri{\`e}re}, \& {Han}}]{Men08}
{Men}, H., {Ferri{\`e}re}, K., \& {Han}, J.~L. 2008, \aap, 486, 819.
  \eprint{arXiv:0805.3454}

\bibitem[{{Mitra} et~al.(2003){Mitra}, {Wielebinski}, {Kramer}, \&
  {Jessner}}]{mitra}
{Mitra}, D., {Wielebinski}, R., {Kramer}, M., \& {Jessner}, A. 2003, \aap, 398,
  993

\bibitem[{{Nota} \& {Katgert}(2010)}]{Katgert}
{Nota}, T., \& {Katgert}, P. 2010, \aap, 513, A65+

\bibitem[{{Noutsos} et~al.(2008){Noutsos}, {Johnston}, {Kramer}, \&
  {Karastergiou}}]{Noutsos}
{Noutsos}, A., {Johnston}, S., {Kramer}, M., \& {Karastergiou}, A. 2008,
  \mnras, 386, 1881. \eprint{0803.0677}

\bibitem[{{Patrikeev} et~al.(2006){Patrikeev}, {Fletcher}, {Stepanov}, {Beck},
  {Berkhuijsen}, {Frick}, \& {Horellou}}]{Fletcher}
{Patrikeev}, I., {Fletcher}, A., {Stepanov}, R., {Beck}, R., {Berkhuijsen},
  E.~M., {Frick}, P., \& {Horellou}, C. 2006, \aap, 458, 441.
  \eprint{arXiv:astro-ph/0609787}

\bibitem[{{Sun} et~al.(2008){Sun}, {Reich}, {Waelkens}, \&
  {En{\ss}lin}}]{Sun08}
{Sun}, X.~H., {Reich}, W., {Waelkens}, A., \& {En{\ss}lin}, T.~A. 2008, \aap,
  477, 573. \eprint{0711.1572}

\bibitem[{{Taylor} et~al.(2003){Taylor}, {Gibson}, {Peracaula}, {Martin},
  {Landecker}, {Brunt}, {Dewdney}, {Dougherty}, {Gray}, {Higgs}, {Kerton},
  {Knee}, {Kothes}, {Purton}, {Uyaniker}, {Wallace}, {Willis}, \&
  {Durand}}]{CGPS}
{Taylor}, A.~R., {Gibson}, S.~J., {Peracaula}, M., {Martin}, P.~G.,
  {Landecker}, T.~L., {Brunt}, C.~M., {Dewdney}, P.~E., {Dougherty}, S.~M.,
  {Gray}, A.~D., {Higgs}, L.~A., {Kerton}, C.~R., {Knee}, L.~B.~G., {Kothes},
  R., {Purton}, C.~R., {Uyaniker}, B., {Wallace}, B.~J., {Willis}, A.~G., \&
  {Durand}, D. 2003, \aj, 125, 3145

\bibitem[{{Taylor} et~al.(1993){Taylor}, {Manchester}, \& {Lyne}}]{pulsars}
{Taylor}, J.~H., {Manchester}, R.~N., \& {Lyne}, A.~G. 1993, \apjs, 88, 529

\bibitem[{{Vall{\'e}e}(2008)}]{Vallee08}
{Vall{\'e}e}, J.~P. 2008, \apj, 681, 303

\bibitem[{{Van Eck} et~al.(2010){Van Eck}, {Brown}, {Stil}, {Rae}, {Mao},
  {Gaensler}, {Shukurov}, {Taylor}, {Haverkorn}, {Kronberg}, \&
  {McClure-Griffiths}}]{VanEck10}
{Van Eck}, C., {Brown}, J.~C., {Stil}, J.~M., {Rae}, K., {Mao}, S.~A.,
  {Gaensler}, B.~M., {Shukurov}, A., {Taylor}, A.~R., {Haverkorn}, M.,
  {Kronberg}, P.~P., \& {McClure-Griffiths}, N.~M. 2010, \apj, submitted

\bibitem[{{Weisberg} et~al.(2004){Weisberg}, {Cordes}, {Kuan}, {Devine},
  {Green}, \& {Backer}}]{Weisberg04}
{Weisberg}, J.~M., {Cordes}, J.~M., {Kuan}, B., {Devine}, K.~E., {Green},
  J.~T., \& {Backer}, D.~C. 2004, \apjs, 150, 317.
  \eprint{arXiv:astro-ph/0310073}

\end{thebibliography}

\end{document}